# Internet of Fly Things For Post-Disaster Recovery Based on Multi-environment

Abdu Saif [1], kaharudin bin dimyati [1*]; Kamarul Ariffin Bin Noordin [1*], Nor Shahida Mohd Shah[2], Qazwan Abdullah,[2] Fadhil Mukhlif [1] and Mahathir Mohamad[3]

[1]Faculty of Engineering, Department of Electrical Engineering, University of Malaya,50603, Kuala Lumpur, Malaysia

[2]Faculty of Engineering Technology, Universiti Tun Hussein Onn Malaysia, Pagoh, Muar, Johor, Malaysia.

[3]Faculty of Applied Science and Technology, Universiti Tun Hussein Onn Malaysia, Pagoh, Muar, Johor, Malaysia

*Abstract*— Natural disasters such as floods and earthquakes immensely impact the telecommunication network infrastructure, leading to the malfunctioning and interruption of wireless services. Consequently, the user devices under the disaster zone are unable to access the cellular base stations. Wireless coverage on an unmanned aerial vehicle (UAV) is considered for providing coverage service to ground user devices in disaster events. This work evaluated the efficient performance of wireless coverage services of UAVs to provide the internet to fly things to help recover the communications link in a natural disaster in multi environments. The results demonstrate the line of sight, non-line of sight, path loss, and coverage probability for the radio propagation environment scenario. Therefore, the path loss and coverage probability are affected by the user devices' elevation angle and distance in the multi-environment system. The user position's optimum user device distance and elevation angle are also investigated to improve the coverage probability, which could be especially useful for the UAV deployment design.

I. INTRODUCTION

The fifth-generation (5G) is the foundation of the internet of things, which is an exponential improvement from the four-generation (4G) counterpart [1]. New technologies are emerging as the foundation of 5G, such as millimeter-wave, which offers a massive bandwidth. More than 1 Gbps of data transmission can be transmitted over many kilometers of distance using a combination of millimeter-wave (mmW) and Free-space optical communication (FSO) [2]. The role of UAVs and D2D communications is to ensure communication in emergency and disaster situations such as earthquakes, floods, and tsunamis, which occur from time to time in many places around the world [3],[4]. The internet portfolio of technologies uses IoT to enhance integration between device-oriented sensor networks and data-oriented applications [5]. One particular criticism of the literature on UAVs and D2D communications is that the standardization activity handles disaster-resilient communication[6], [7]. A great deal of previous work on UAVs and D2D communications focused on IoT platforms to enable novel solutions for all things that have a significant qualitative change on people's lives, such as smart cities, grids, homes, and connected vehicles [6], [8]. Overall, these studies highlighted the need for IoTs as a promising paradigm that is rapidly gaining ground in future wireless communications such as 5G [9, 10].

UAVs are considered crucial for developing futures generations in wireless networks[11]. Therefore, the most common goal of researchers attracts quick development for the application of UAVs that provide wireless services to society, industry, and the government in different environmental scenarios[12]. Further work on UAVs is required to establish viability, remote surveillance, monitoring, relief operations, package delivery, and communication backhaul infrastructure based on Massive MIMO System [2].

Besides, a large volume of research that includes accurate air-to-ground (A-G) propagation channel models[13]. The design and evaluation of UAVs, communication links for control / non-payload, as well as payload data transmissions would be needed to build a complete image of UAV scenarios and increase system throughput[14]. Meanwhile, the primary challenge faced by many researchers is the physical damage done to the network infrastructure or power outages as the aftermath of a natural disaster[15]. UAVs may serve as flying base stations to meet the connectivity needs of individuals in disaster-affected areas[16],[17]. The UAV-IoFT can enable communication services while wireless communication networks are dysfunctional during natural disasters[8]. UAV-IoFT can fly in any condition to provide a large coverage area with reduced power consumption and secure people in need of help during a natural disaster [18].

A. Contribution

The UAV-IoFT communications increase the system capacity and spectrum efficiency under control by the cellular system. In any natural disaster event, public safety plays a vital role in communication recovery and saving lives. Therefore, this work focuses on providing alternative coverage services access to user devices in infrastructure damage due to natural disaster scenarios. Furthermore, it deploys the UAV-IoFT to provide wireless coverage services in necessity when ground user devices cannot obtain coverage from the ground base station. The key contributions include:

- After a disaster, to boost wireless coverage services, including a dual-slope Los / NLoS propagation model for both route loss and small-scale fading.
- To provide reliable connectivity ubiquitous coverage probability, including the effect of user device distance from the UAV for various propagation environments. The Los/NLoS communication link, path loss, and coverage probability are based on the ratio of system sum-rate to the total power consumption, and the bits/Joule will be enhanced.

## II. SYSTEM MODEL

A scenario where a distributed ground user device with a D2D communications architecture is considered. The user devices coexist with a UAV Communication system in the same frequency band for sightline and non-sightline communication links in the disaster zone, such as a spectrum sharing system.

Fig 1. Shows three types of ray signals at the receivers of ground user devices, including direct, reflected, and diffracted. The direct rays between the UAV and ground user devices represent signals that travel freely. However, the ray signals between the UAV and the ground user device by the diffracted or reflected represent indirect links due to obstacles' interaction.

The main goal of this proposed model is to verify the reliability and availability of QoS signals at ground user device receivers during a natural disaster. The user devices can obtain the UAV-IoFT service and hotspot service to all devices inside the disaster zone through the spectrum sharing system. Hence, the Public Safety Internet of Fly Things (PS-IoFT) acts to minimize the transmission power of each ground user device (GUD), D2D communications, and pre-defined SINR requirements. Hence, the UAV-IoFT can fly autonomously and provide wireless coverage services with serval applications in natural disaster communication scenarios. Furthermore, UAV-IoFT was integrated with (PS-IoFT) as a promising technology to support the fifth-generation (5G) wireless communications. Those technologies can change location dynamically to respond to an emergency and have fast reconfiguration to provide effective communication and quicker disaster recovery.

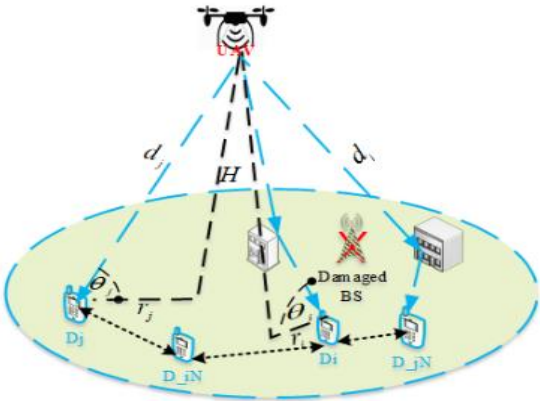

Fig. 1. Illustration of UAV providing the Internet of Fly Things (IoFT)

### A. Channel Model

The direct line of sight channel between UAV and ground user devices through the air to ground channels and indirect non-line of sight between ground users and building are provided. UAV was deployed to provide IoFT wireless networks with an improved probability of serving ground user devices via a Los link, which has lower attenuation of propagation than an NLoS link. Besides, due to the blocking effect caused by construction, trees, etc., it is generally unavoidable that the connexion between UAV to $D_i$ and $D_j$ is an NLoS connexion. The optimal UAV altitude considers maximizing the coverage probability and reliable connectivity in disaster situations. The UAV altitude is h, and the reduction of coverage is $r_0$, hence the UAV-GUDs distance denoted as follows:

$$d = \sqrt{r_o^2 + h^2} \qquad (1)$$

The possibility of a Los and an NLoS link between the UAV and GUDs are denoted by $P_{Los}(\varphi)$ and $P_{NLos}(\varphi)$. This is shown as follows [19]:

$$P_{Los}(r_0) = \frac{1}{1 + a\,exp(-b(\varphi(r_0,h)-a))} \qquad (2)$$

$$P_{NLoS}(r_0) = 1 - P_{Los}(r_0) \qquad (3)$$

where $\varphi$ is the elevation angle of the GUDs $\in D_i, D_j$, and $a$ and $b$ are parameters that affect the S-curve parameters that vary according to the environment, such as urban, suburban, dense urban, and high-rise urban. Furthermore, from Eq (2), the link is more likely to be an LoS communication link with a more significant elevation angle.

Fig 4. shows that user devices are uniformly distributed within a geographical area of size 1 km x 1 km. The UAV represents the backhaul link for the set of ground user devices in the pot disaster. The UAV can provide full coverage to user devices through Los/NLoS Links.

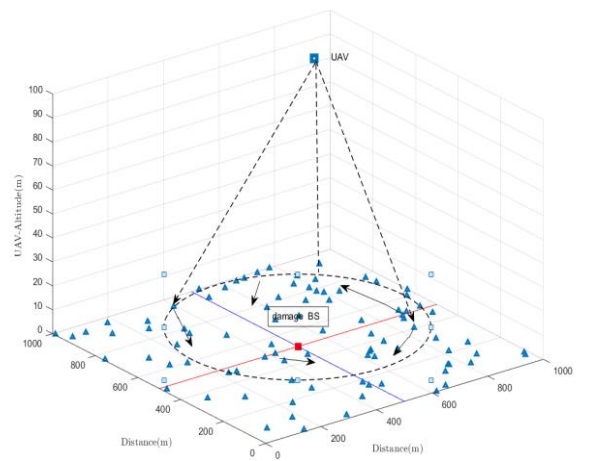

Fig. 2. Illustration location of UAV, GUDs

### B. Analysis of Path Loss

Path loss propagation is a critical factor that affects the wireless channel, including the attenuation of radiated signals with distance, user device elevation angle, and the distance from the UAV. Therefore, the pathless ground user devices include Los and NLoS links with multipath fading and shadowing from the transmitted signals due to blocking and large-scale path loss obstacles. The average path loss of ground user's device $D_i$ and $D_j$ UAV can be calculated as:

$$PL(dB) = FSP + AL \quad (4)$$

where $FSP = 20 \log(\frac{4\pi f_c d(i,j)}{c})$ is the free space path loss, $AL = \mu_{LoS} PLoS + \mu_{NLoS} PNLoS$ is the average additional loss to the free space loss, $f_c$ denotes as carrier frequency, $d_{(i,j)}$ denotes as the distance from the UAV to $i^{th}, j^{th}$ GUDs, and $c$: represents lighting speed.

*C. Coverage Probability of Downlink*

In the downlink coverage, we consider the shadowing effect in LoS and NLoS communication links. Hence, the communication link between the UAV and GUDs is affected by different Gaussian distributions as follows:

$$\varphi_{LoS} \sim N(\mu_{LoS}, \sigma_{LoS}^2) \; \varphi_{NLoS} \sim N(\mu_{NLoS}, \sigma_{NLoS}^2)$$

where $\sigma^2$ is the noise power and $\mu$ is the additional path loss for Los and NLoS links, respectively. Therefore, the coverage probability is defined as the probability that the received power is more significant than a specified threshold. Then, $P_{cov} = P(p_r \geq p_{min})$
where $p_r$ is the received signal power at the ground user device. The coverage probability in the disaster area is given by Eq (5). When the ground users device are $\in D_i, D_j$, the condition coverage probability achieved by the ground user devices distance can be expressed as follows:

$$P_{cov} = P_{LoS} Q[A] + P_{NLoS} Q[B] \quad (5)$$

Furthermore:

$$A = \frac{p_{min} + PL(dB) - p_t - GdB + \mu_{LoS}}{\sigma_{LoS}^2}, \quad B = \frac{p_{min} + PL(dB) - p_t - GdB + \mu_{NLoS}}{\sigma_{NLoS}^2}$$

where $PL(dB)$ represents the path loss, $GdB = 3dB$ represents the antenna gain, and $Q$ is the function of the Los and NLoS communication links connection between the ground user devices and the UAV, respectively.

TABLE I.  SIMULATION PARAMETERS

| ENVIRONMENT PARAMETERS [20] | | | | |
|---|---|---|---|---|
| Parameters | a | b | $\mu_{LoS}$ | $\mu_{NLoS}$ |
| Suburban | 5.2 | 0.35 | 0.1 | 21 |
| urban | 10.6 | 0.18 | 1 | 20 |
| dense urban | 11.95 | 0.14 | 1.6 | 23 |
| highrise urban | 26.5 | 0.13 | 2.3 | 34 |

### III. NUMERICAL RESULTS AND ANALYSIS

In this section, the simulation results are presented to analyze the performance of IoFT communication systems with UAV to recover disaster communications. The parameters of the simulation can be shown as follows. The channel bandwidth is 5 MHz, while the additive white Gaussian noise (AWGN) power for Los and NLoS is $\sigma = -174 dBm$. The $p_{UAV}$ represents the UAV transmit Power of 10W, while $D_i$, $D_j$ are the transmit power of the ground user devices of 1 W. The IoFT network radius is between 500 to 1000 m, with a UAV providing the IoFT access. The environment parameters in Table (I) are represented by different communication environments such as suburban, urban, dense urban, and high-rise urban. The received wireless signals from the UAV to the ground user devices in the downlink are usually modelled by probability LoS/NLoS, path loss, and coverage probability for the selection environments versus the user devices elevation angles and user devices distance. Fig 3. shows that the performance of LoS probability increased when the user device's elevation angles risen simultaneously for the same level of coverage across all network environments. It can also be seen that the maximum probability of LoS is achieved at the elevation angle of 20º in sub-urban and of 45º in urban, dense urban, and high-rise urban. Thus, UAVs can increase the gain and fly over a region and operate optimally within the receiver's LoS range. Conversely, Fig 4. shows that the performance of NLoS probability decreased when the user devices' angle increases across different environment scenarios. The minimum possibility of NLoS was achieved at the elevation angle of 20º in sub-urban and 45o in urban, dense urban, and high-rise urban, due to the large scale of NLoS affected by the density of buildings and path loss obstacles.

Fig 5. shows that the path loss was affected by the ground user devices' elevation angles in the targeted environment. The path loss was from 76 dB to 110 dB, and the elevation angles increased from 0 to 15 m for both targeted environment scenarios due to the power consumption by ground user devices. Therefore, the path loss decreased from maximum to 94 dB and 84 dB, and the user device's distance increased from 100 m to 250 m for the high-rise urban environment. However, the path loss still increased from 76 dB to 101 dB in urban areas when the user device distance increased from 15m to 300m. Furthermore, the path loss dropped from 101 dB to 100 dB when the user device distance exceeded 300m. The path loss increased from 76 dB to 106 dB, and 110 dB for dens-urban and high rise urban, when the user device distance increased from 15m to 500m. Finally, a lower path loss and small-scale fading were found in the A-G channel compared to a G-G link while having a longer link length that deteriorates the received SNR. However, in the suburban area, the path loss increased from 82.5 dB to 86dB, while the distance increased from 270 m to 500 m. This for a single model city and cannot be widely generalized for different urban environments.

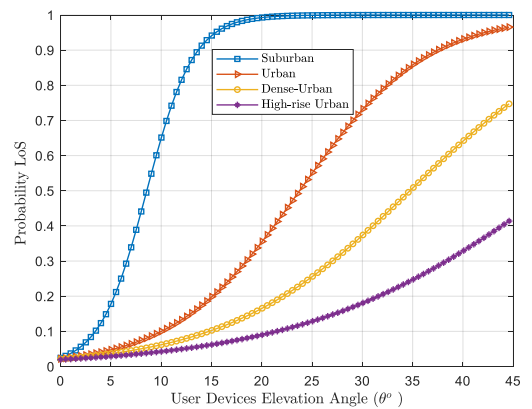

Fig.3. Illustration of probability LoS Versus the User devices Elevation Angle.

Fig 3. shows that the coverage probabilities for all environment scenarios improve when the user device's elevation angle is increased due to the UAV expanding the coverage area for user devices nearest to UAV and increase the Los with sufficient SINR. Therefore, the high-rise urban had a higher coverage probability versus the user devices' elevation angle due to increasing the NLoS SNR at the distribution nodes for a high density of structures in man-made environments. However, the urban, dense urban, and suburban area coverage probabilities increased. Simultaneously, the user devices' elevation angles that affect signals are more vulnerable to the increasing number of LoS interfering at the receivers' destination nodes. The reason is that UAVs altitude can gain height and hover over a particular region and operate within the line of sight (LoS) range at the receiver.

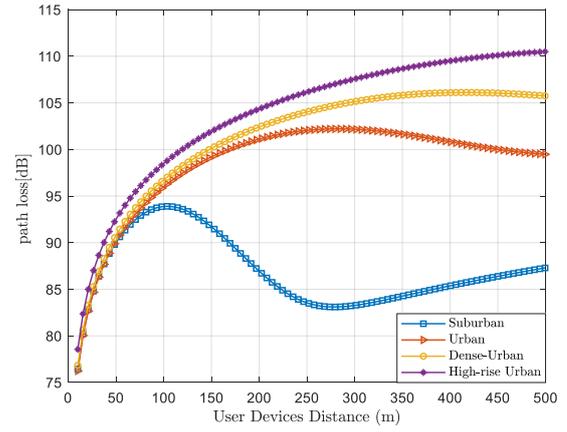

Fig 5. Illustration analysis Path Loss versus user devices distance.

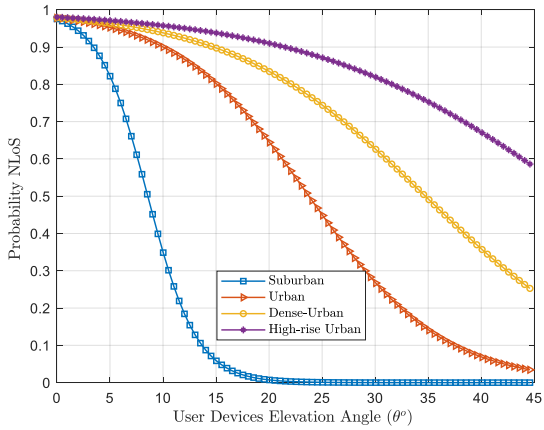

Fig. 4. Illustration probability NLoS versus User devices Elevation Angle.

Fig 6. shows the performance of the coverage probability versus the user device's distance. It can be seen that the high-rise urban achieved maximum coverage probability due to the increasing Los and NLoS SNR at the receiver nodes. However, the low coverage probability is found in the dense urban, urban, and suburban. This is due to the received single only from the LoS at the destination nodes and slow Los for those environment scenarios. Therefore, the coverage probability increased at the user devices distance to less than 100m, and decreased at the range of user devices distance between the 100m to 500m, due to the large scale path loss that occurred. The user device distance increased as the SNRR increased, and this probability continues to improve steadily. The behavior of the Los and NLoS probability is affected by the user device's distance. Due to the range of coverage area, the user devices are close enough that no LoS-blocking buildings are affected by the received signals in GUDs, which is the reason for guaranteeing the stronger Los signals.

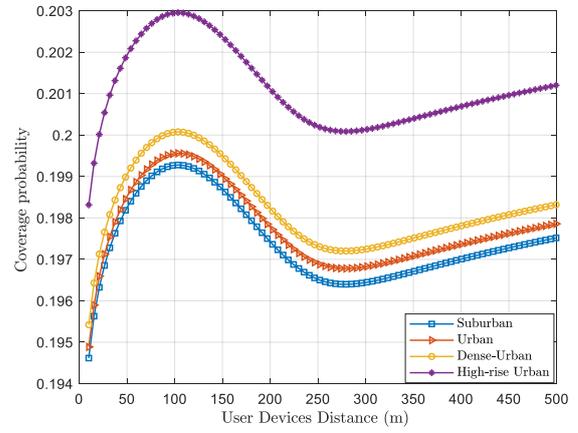

Figure 6. Illustration analysis Coverage probability versus user devices distance.

## IV. Conclusions

This paper identified a proposed UAV to provide emergency coverage services in disaster areas and visibility of the communications between the in-coverage and out-of-coverage areas. The performance of coverage probability varies based on the selected environment scenarios. Therefore, it is crucial to consider UAV network parameters such as elevation angle and user devices distance and environmental types such as suburban, urban, dense urban, and high rise urban. The performance of the UAV network's coverage probability was affected by the environmental parameters as a function of the attitudes and user device's distance.

## V. Future work

The researcher recommends using energy harvesting techniques with clustering and multi-hop D2D communications to prolong the lifetime of the energy network and ensure reliable connectivity during disaster phases. Besides, integrating the UAV wireless coverage services with smart devices compatible with the Internet of fly things (IoFT) and the Internet of public safety things (IoPST) also merits further investigation.

ACKNOWLEDGMENT

This work is supported by the University of Malaya, Faculty of Electrical Engineering under the DARE project (Grand ID: IF035A-2017 & IF035-2017) and also supported by the Ministry of Higher Education Malaysia under the Fundamental Research Grant Scheme (FRGS) Vot FRGS/1/2018/TK04/UTHM/02/14 and also partially sponsored by Universiti Tun Hussein Onn Malaysia under TIER1 Grant (Vot H158)

REFERENCES

[1] Q. Abdullah *et al.*, "A Brief Survey And Investigation Of Hybrid Beamforming For Millimeter Waves In 5G Massive MIMO Systems," *Solid State Technology,* vol. 63, no. 1s, pp. 1699-1710, 2020.

[2] A. Salh, N. S. M. Shah, L. Audah, Q. Abdullah, W. A. Jabbar, and M. Mohamad, "Energy-efficient power allocation and joint user association in multiuser-downlink massive MIMO system," *IEEE Access,* vol. 8, pp. 1314-1326, 2019.

[3] A. Saif, K. Dimyati, K. A. Noordin, N. S. M. Shah, Q. Abdullah, and F. Mukhlif, "Unmanned Aerial Vehicles for Post-Disaster Communication Networks," in *2020 IEEE 10th International Conference on System Engineering and Technology (ICSET)*, 2020: IEEE, pp. 273-277.

[4] S. Alsamhi *et al.*, "Green Internet of Things using UAVs in B5G Networks: A Review of Applications and Strategies," *Ad Hoc Networks,* p. 102505, 2021.

[5] G. Deepak, A. Ladas, Y. A. Sambo, H. Pervaiz, C. Politis, and M. A. Imran, "An overview of post-disaster emergency communication systems in the future networks," *IEEE Wireless Communications,* vol. 26, no. 6, pp. 132-139, 2019.

[6] A. Saif *et al.*, "Unmanned Aerial Vehicle and Optimal Relay for Extending Coverage in Post-Disaster Scenarios," *arXiv preprint arXiv:2104.06037,* 2021.

[7] Y. Sun, Z. Ding, and X. Dai, "A User-Centric Cooperative Scheme for UAV-Assisted Wireless Networks in Malfunction Areas," *IEEE Trans. Commun.,* vol. 67, no. 12, pp. 8786-8800, 2019.

[8] A. Saif *et al.*, "Distributed Clustering for User Devices Under Unmanned Aerial Vehicle Coverage Area during Disaster Recovery," *arXiv preprint arXiv:2103.07931,* 2021.

[9] J. Lu, S. Wan, X. Chen, and P. Fan, "Energy-efficient 3D UAV-BS placement versus mobile users' density and circuit power," in *2017 IEEE Globecom Workshops (GC Wkshps)*, 2017: IEEE, pp. 1-6.

[10] F. Mukhlif, K. A. B. Nooridin, Y. A. AL-Gumaei, and A. S. AL-Rassas, "Energy harvesting for efficient 5g networks," in *2018 International Conference on Smart Computing and Electronic Enterprise (ICSCEE)*, 2018: IEEE, pp. 1-5.

[11] A. Gupta, S. Sundhan, S. K. Gupta, S. Alsamhi, and M. Rashid, "Collaboration of UAV and HetNet for better QoS: a comparative study," *International Journal of Vehicle Information and Communication Systems,* vol. 5, no. 3, pp. 309-333, 2020.

[12] S. H. Alsamhi, O. Ma, M. S. Ansari, and S. K. Gupta, "Collaboration of drone and internet of public safety things in smart cities: An overview of qos and network performance optimization," *Drones,* vol. 3, no. 1, p. 13, 2019.

[13] W. Khawaja, I. Guvenc, D. W. Matolak, U.-C. Fiebig, and N. Schneckenburger, "A survey of air-to-ground propagation channel modeling for unmanned aerial vehicles," *IEEE Communications Surveys & Tutorials,* vol. 21, no. 3, pp. 2361-2391, 2019.

[14] Q. Abdullah *et al.*, "Maximising system throughput in wireless powered sub-6 GHz and millimetre-wave 5G heterogeneous networks," *Telkomnika,* vol. 18, no. 3, pp. 1185-1194, 2020.

[15] S. H. Alsamhi, O. Ma, M. S. Ansari, and F. A. Almalki, "Survey on collaborative smart drones and internet of things for improving smartness of smart cities," *IEEE Access,* vol. 7, pp. 128125-128152, 2019.

[16] S. Alsamhi, F. Almalki, O. Ma, M. Ansari, and M. Angelides, "Performance optimization of tethered balloon technology for public safety and emergency communications," *Telecommunication Systems,* pp. 1-10, 2019.

[17] A. Khan *et al.*, "RDSP: Rapidly Deployable Wireless Ad Hoc System for Post-Disaster Management," *Sensors,* vol. 20, no. 2, p. 548, 2020.

[18] J. Kakar, A. Chaaban, V. Marojevic, and A. Sezgin, "UAV-aided multi-way communications," in *2018 IEEE 29th Annual International Symposium on Personal, Indoor and Mobile Radio Communications (PIMRC)*, 2018: IEEE, pp. 1169-1173.

[19] M.-N. Nguyen, L. D. Nguyen, T. Q. Duong, and H. D. Tuan, "Real-time optimal resource allocation for embedded UAV communication systems," *IEEE Wirel. Commun. Lett.,* vol. 8, no. 1, pp. 225-228, 2018.

[20] B. Wang, J. Ouyang, W.-P. Zhu, and M. Lin, "Optimal Altitude of UAV-BS for Minimum Boundary Outage Probability with Imperfect Channel State Information," in *2019 IEEE/CIC International Conference on Communications in China (ICCC)*, 2019: IEEE, pp. 607-611.